\documentclass[twocolumn,showpacs,preprintnumbers,amsmath,amssymb,showpacs,aps,prb,lengthcheck]{revtex4-1}
\usepackage{amssymb}
\usepackage{graphicx}
\usepackage{dcolumn}
\usepackage{bm}
\usepackage{hyperref}
\usepackage{makeidx}
\usepackage[english]{babel} 
\usepackage[latin1]{inputenc}
\usepackage{color}
\usepackage{graphicx}
\usepackage{amsmath}
\usepackage{latexsym}
\usepackage{amsthm}

\makeindex
\def\>{\right\rangle}
\def\<{\left\langle}
\def\be{\begin{equation}}
\def\ee{\end{equation}}
\def\ba{\begin{array}{l}}
\def\ea{\end{array}}

\def\beq{\begin{eqnarray}}
\def\eeq{\end{eqnarray}} 

\begin{document}

\title{Generating and controlling spin-polarized currents induced by a quantum spin Hall antidot}

\author{G. Dolcetto$^{1,2,3}$, F. Cavaliere$^{1,2}$, D. Ferraro$^{2,3,4}$, M. Sassetti$^{1, 2}$}
 \affiliation{$^1$ Dipartimento di Fisica, Universit\` a di Genova,Via Dodecaneso 33, 16146, Genova, Italy.\\
$^2$ CNR-SPIN, Via Dodecaneso 33, 16146, Genova, Italy.\\ 
$^3$ INFN, Via Dodecaneso 33, 16146, Genova, Italy.\\
$^4$ Universit\'{e} de Lyon, F\' ed\' eration de Physique Andr\' e Marie Amp\` ere, CNRS - Laboratoire de Physique de l'Ecole Normale Sup\' erieure de Lyon, 46 All\' ee d'Italie, 69364 Lyon Cedex 07, France.}

\date{\today}
\begin{abstract}
\noindent
We study an electrically-controlled quantum spin Hall antidot embedded in a two-dimensional topological insulating bar. Helical edge states around the antidot and along the edges of the bar are tunnel-coupled. The close connection between spin and chirality, typical of helical systems, allows to generate a {\em spin-polarized} current flowing across the bar. This current is studied as a function of the external voltages, by varying the asymmetry between the barriers. For asymmetric setups, a switching behavior of the spin current is observed as the bias is increased, both in the absence and in the presence of electron interactions.
This device allows to generate and control the spin-polarized current by simple electrical means.

\end{abstract}

\maketitle

\section{Introduction}
Two dimensional topological insulators,~\cite{Hasan10, Qi11} showing quantum spin Hall (QSH) effect,~\cite{Kane05a, Kane05b} were recently subjected to intensive theoretical and experimental research.
The theoretical prediction by Bernevig \textit{et al.}~\cite{Bernevig06b}, about the possibility to realize the QSH effect in Hg-Te quantum wells, was experimentally proved by Konig \textit{et al.}~\cite{Konig07}, who showed that these systems are characterized by gapless edge states at zero magnetic field.
The absence of time reversal-breaking mechanism ensures the robustness of these edge states from backscattering.~\cite{Bernevig06b, Roth09}
The main property of QSH effect is that the edge states are helical,~\cite{Wu06} with locking between spin and momentum degrees of freedom.\\
Several theoretical proposal were made in order to confirm the connection between spin and chirality.
In particular, quantum point contacts (QPCs),~\cite{Strom09, Teo09, Hou09, Liu11, Schmidt11} extended contacts,~\cite{Dolcetto12} interferometric setups,~\cite{Dolcini11, Virtanen11, Romeo12, Ferraro12} as well as scanning-tunneling microscopy~\cite{Das11, Konig12} turned out to be appropriated for this task.
Recently, quantum dot realized in Hg-Te quantum wells have been theoretically studied.~\cite{Chang11, Tkachov11, Chu09, Timm11, Posske13}
The fast development in the experimental techniques will soon allow to verify these theoretical proposals.\\
The important role played by the spin degrees of freedom makes these system fascinating for future applications.
The ability to generate and control spin currents would open the possibility to realize promising devices for the development of spintronics.~\cite{Sukhanov12}
Recent proposals of devices for spin manipulaton, based on two-dimensional topological insulators, rely on Aharonov-Bohm and Fabry P\'{e}rot interferometers,~\cite{Dolcini11, Maciejko10, Citro11} the application of a magnetic field at a pn junction~\cite{Akhmerov09} and tunable wavefunction overlapping in nanoconstrictions.~\cite{Romeo12, Krueckl11}
However, simple two-terminal configurations are often not able to induce a global spin-polarized current flowing through the system, unless magnetic fields or ferromagnets are present.~\cite{Maciejko10, Timm11}
The aim of this paper is to analyze the possibility of generating and controlling spin-polarized current in a two-terminal QSH antidot system.
This geometry was already proposed for detecting peculiar properties of fractional charges in fraction quantum Hall bar.~\cite{Merlo06}
The interest related to this device is that the spin-polarized current can be controlled only by electrical means, thus overcoming the need for magnetic materials.\\
The antidot is embedded in a QSH bar and coupled, via tunnel barriers, to the edges of the bar.
We will show that this setup, by virtue of the close link between spin and chirality, allows not only to generate, but also to control the spin-polarized current along the edges of the bar in a very simple and efficient way, both in the absence and in the presence of electron interactions.
We will show that the asymmetry between the tunnel barriers is a key ingredient in order to produce spin-polarized current.\\
The paper is organized as follows.
In Sec. \ref{model} we introduce the theoretical model for the device.
In Sec. \ref{master equation} we describe the master equation approach adopted in order to calculate the transport properties and we define the relations between the current through the antidot and along the edges of the QSH bar.
Section \ref{spin-polarized current} contains the main results of our research. The spin-polarized current is analyzed as a function of the external voltages, by varying the asymmetry between the barriers and the strength of the electron interactions.
Section \ref{conclusions} is devoted to the conclusions.

\section{Model}\label{model}

\begin{figure}[!ht]
\centering
\includegraphics[scale=0.33]{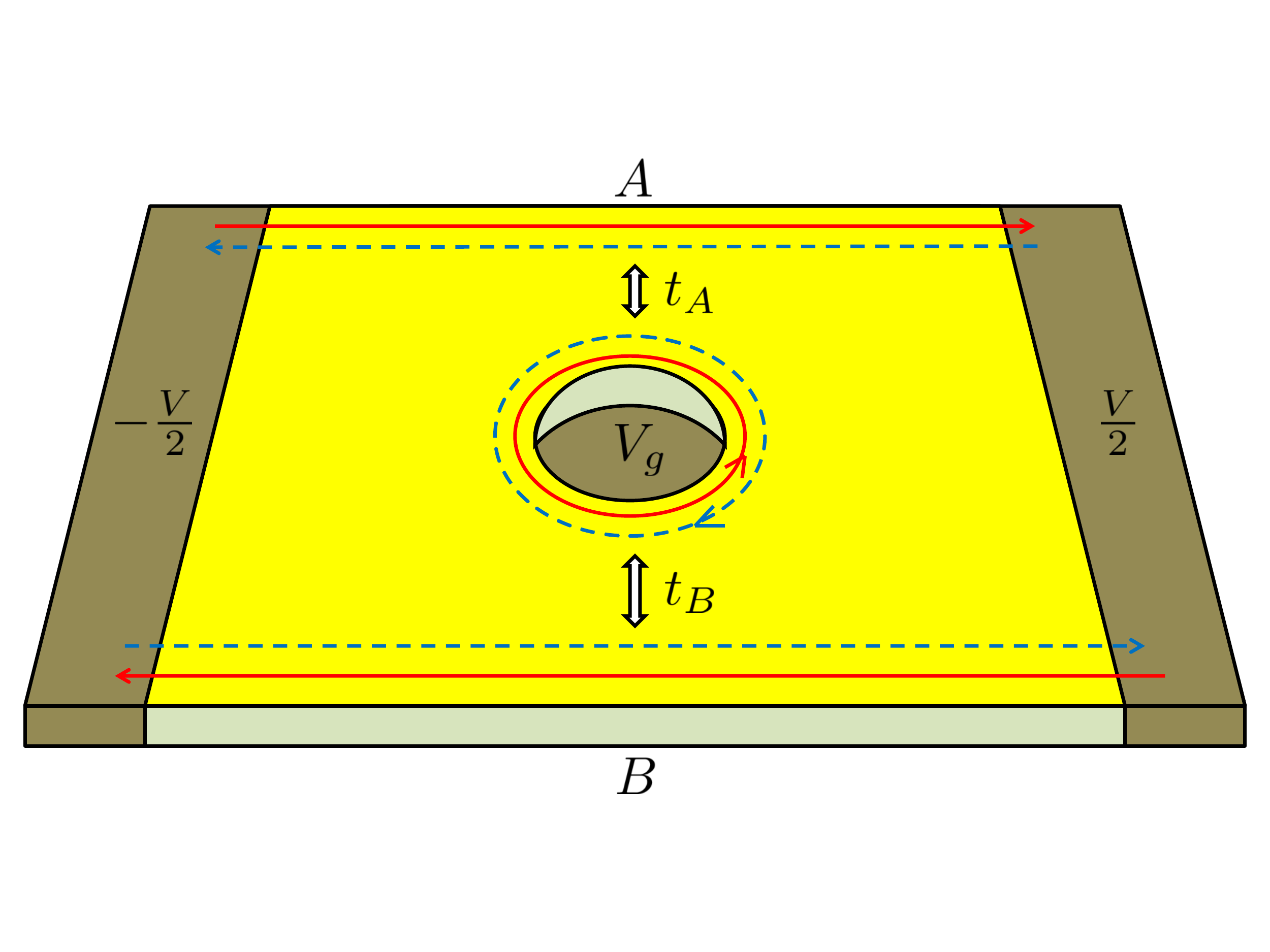}
\caption{(Color online) Schematic representation of an antidot embedded in a QSH bar. The edges of the bar are biased in a two-terminal configuration. Helical edge states appear at the boundaries between the QSH phase and the vacuum.
Full red (dashed blue) lines represent spin up (down) electrons.}\label{antidot}
\end{figure}

The system is schematically shown in Fig. \ref{antidot}. An antidot is created inside a QSH bar in a two terminal setup. Gapless helical edge states appear along the edges of the bar and around the antidot. We model the coupling between the antidot ($ad$) and the upper ($A$) and lower ($B$) edges by tunneling barriers.
The system is driven out of equilibrium by applying a bias potential $V$. The chemical potential in the antidot region can be tuned through a gate voltage $V_g$. The full Hamiltonian of the system
\begin{equation}\label{H_tot}
H=H_{A}+H_{B}+H_{ad}+H^{(A)}_{t}+H^{(B)}_{t}+H_{V}
\end{equation}
consists of the sum of the leads ($H_A+H_B$) and the antidot ($H_{ad}$) uncoupled Hamiltonians, the tunneling barriers ($H^{(A)}_{t}+H^{(B)}_{t}$) and the coupling of the system with the external circuit ($H_{V}$).

The helical edge states around the antidot are described in terms of helical Luttinger liquid (HLL)~\cite{Wu06} ($\hbar=1$)
\begin{eqnarray}\label{Had}
H_{ad}&=&-iv_{F}\int_{0}^{L} dx \left (\psi^{\dagger}_{ad,\uparrow}\partial_x\psi_{ad,\uparrow}-\psi^{\dagger}_{ad,\downarrow}\partial_x\psi_{ad,\downarrow}\right ) \nonumber \\
&+&\frac{g}{2}\int_{0}^{L} dx\left (\sum_{\sigma=\uparrow,\downarrow}\psi^{\dagger}_{ad,\sigma}\psi_{ad,\sigma}\right )^2
,\end{eqnarray}
where $v_F$ is the Fermi velocity, $L$ is the circumference of the antidot and $g$ is proportional to the screened Coulomb repulsion.
The bosonization technique~\cite{Giamarchi03} allows to write the electron operator
\begin{equation}\label{bos}
\psi_{ad,\sigma}(x)=e^{\pm ik_F x}\frac{\mathcal{F}_{\sigma}}{\sqrt{2\pi a}}e^{\pm 2\pi i\frac{N_{\sigma}}{L}x} e^{-i\sqrt{2\pi} \phi_{\sigma}(x)}
\end{equation}
in terms of the bosonic field $\phi_{\sigma}(x)$, where the sign $+$ ($-$) is for spin up $\sigma=\uparrow$ (spin down $\sigma=\downarrow$) electrons.
The chiral index is omitted, since it is directly linked to the spin.
In Eq. ~\eqref{bos}, $\mathcal{F}_{\sigma}$ represents a Klein factor, $n_{\sigma}$ is the excess number of electrons with respect to a ground state number $n_{0,\sigma}$, $a$ is a short length cut-off and $k_F=\pi (n_{0,\uparrow}+n_{0,\downarrow})/L$ is the Fermi momentum. The Hamiltonian ~\eqref{Had} can be diagonalized and reads
\begin{equation}\label{H_diag}
H_{ad}=\sum_{\nu=1,2}\sum_{q>0}\epsilon_{\nu}(q) d^{\dagger}_{q,\nu}d_{q,\nu}+\frac{E_n}{2}\left (n-n_{\rm{g}}\right )^2+\frac{E_s}{2}s^2.
\end{equation}
Here, $d_{q,\nu}$ are annihilation operators of the collective plasmon excitations, while $n=n_{\uparrow}+n_{\downarrow}$ and $s=n_{\uparrow}-n_{\downarrow}$ represent the total excess charge and spin (a part for dimensional factors $-e$ and $\hbar/2$) with respect to the ground state value $n_0$ and $s_0$.
The capacitive coupling between the antidot and the gate is taken into account through the term in $n_{\rm{g}}=V_gC_g/e$.~\cite{CavalierePRL04}
Due to boundary conditions, the wave number is quantized, $q=2\pi m_{\nu}/L$ and $\epsilon_{\nu}(q)=m_{\nu}\epsilon_0/K$, with $\epsilon_0=2\pi v_F/L$ the level spacing and $m_{\nu}$ positive integers.
Three different energies characterize the plasmon, the charge and the spin excitations respectively
\begin{equation}\label{energies}
\epsilon=\frac{1}{K}\epsilon_{0} \ \ \ \ \ \  E_{n}=\frac{1}{K^2}\frac{\epsilon_{0}}{2} \ \ \ \ \ \ E_{s}=\frac{\epsilon_{0}}{2}.
\end{equation}
The coefficient $K\equiv \left (1+g/(\pi v_F)\right )^{-\frac{1}{2}}$ encodes the strength of the interactions.
For the non interacting case ($g=0, K=1$) one has $\epsilon=2E_n=2E_s$.
Even in the absence of electron interactions, the contributions corresponding to the charging energy ($E_n$) and to the spin addition energy ($E_s$) are finite, due to the discrete nature of the dot and the Pauli principle.
In addition, despite our microscopic model provides quantitative estimates for these energies, several influences that occur in experimental setups are here neglected.
Indeed, long range interaction effects and coupling with gates can lead to an effective charging energy greater than the one obtained in the microscopic model. \cite{Kleimann}
More robust is the spin addition energy, which is affected at most by the exchange part of the interactions.
Therefore, we treat in the following $E_n$ as a free parameter with $E_n\gg E_s$.\\
One of the peculiar features of QSH systems is the absence of spin-charge separation even in presence of electron interactions. Indeed, the plasmon excitations are characterized by a unique energy scale,~\cite{Dolcetto12} in contrast to the usual interacting spinful one dimensional systems.~\cite{CavalierePRL04, Sassetti98}\\
Analogously, the Hamiltonian of the edges $\lambda=A,B$, supposed infinite, is
\begin{eqnarray}\label{H_edges}
H_{\lambda}&=&-iv_{F}\int_{-\infty}^{\infty} dx \left (\psi^{\dagger}_{\lambda,\uparrow}\partial_x\psi_{\lambda,\uparrow}-\psi^{\dagger}_{\lambda,\downarrow}\partial_x\psi_{\lambda,\downarrow}\right ) \nonumber \\
&+&\frac{g}{2}\int_{-\infty}^{\infty} dx\left (\sum_{\sigma=\uparrow,\downarrow}\psi^{\dagger}_{\lambda,\sigma}\psi_{\lambda,\sigma}\right )^2
,\end{eqnarray}
where $\psi_{\lambda,\sigma}(x)$ is the electron field operator in the edge $\lambda$.
Despite the edges and the antidot can be subject to different screening mechanisms, in this work we assume a unique interaction parameter $g$ for the electron interaction.
The application of a bias voltage $V$ in the two terminal setup is described by
\begin{eqnarray}\label{H_c}
H_V&=&\frac{eV}{2}\left (n_{A,\uparrow}+n_{B,\downarrow}-n_{A,\downarrow}-n_{B,\uparrow}\right )\nonumber \\
&=&\frac{eV}{2}\left (s_A-s_B\right )
,\end{eqnarray}
where $n_{\lambda,\sigma}$ is the number of extra charges on edge $\lambda$ and $s_{\lambda}=n_{\lambda,\uparrow}-n_{\lambda,\downarrow}$. The coupling between edges and antidot is described by pointlike tunneling barriers
\begin{equation}\label{H_t}
H_t^{(\lambda)}=t_{\lambda}\sum_{\sigma=\uparrow,\downarrow}\psi_{ad,\sigma}^{\dagger}(x_{\lambda}) \psi_{\lambda,\sigma}(x_0)+h.c.
\end{equation}
with $x_{\lambda}, x_0$ the points where tunneling occurs and $t_{\lambda}$ the tunneling amplitude.
The latter can be controlled through gate voltages placed above the barriers, thus allowing to increase or decrease their opacities.

\section{Tunneling dynamics}\label{master equation}
The transport properties are determined by the tunneling processes between the edges of the bar and the antidot region~\cite{Beenakker91}.
The state of the antidot is specified by the excess charge $n$, excess spin $s$ and distribution $\left \{m_{\nu}\right \}$ of the plasmons.
The corresponding occupation probability is determined by the tunneling rate $\Gamma^{(\lambda)}_{i\to f}$ from an initial state $|i\rangle=\left | n_i,s_i,\left \{m_{\nu,i}\right \}\right \rangle$ to a final one $|f\rangle=\left | n_f,s_f,\left \{m_{\nu,f}\right \}\right \rangle$ across the barrier $\lambda$.
The sequential nature of the tunneling processes restricts the possible final states to $n_f=n_i+\Delta_n$, $s_f=s_i+\Delta_s$, with $\Delta_n,\Delta_s=\pm 1$.
At the lowest order in the tunneling amplitudes, the transition rates can be written in the compact form
\begin{equation}\label{rate}
\Gamma^{(\lambda)}_{i\to f}=|t_{\lambda}|^2\int dt e^{-i\Delta U^{(\lambda)}_{i\to f}t}\mathcal{G}_{ad}(t)\mathcal{G}_{\lambda}(t),
\end{equation}
where $\Delta U^{(\lambda)}_{i\to f}\to\Delta U^{(\lambda)}_{\Delta_n,\Delta_s}(n,s)$ with
\begin{equation}\label{DeltaU}
\Delta U^{(\lambda)}_{\Delta_n,\Delta_s}(n,s)=\mu_{\Delta_n,\Delta_s}(n,s)\mp\frac{eV}{2}\Delta_s
.\end{equation}
The sign $\mp$ in Eq.~\eqref{DeltaU} distinguishes tunneling across barrier A ($-$) and B ($+$).
The quantity $\mu_{\Delta_n,\Delta_s}(n,s)$ is the chemical potential of the antidot without plasmon excitations
\begin{equation}\label{mu}
\mu_{\Delta_n,\Delta_s}(n,s)=E_n\left [\frac{1}{2}+\Delta_n(n-n_{\rm{g}})\right ]
+E_s\left [\frac{1}{2}+\Delta_ss\right ].
\end{equation}
In Eq. ~\eqref{rate} we introduced
\begin{equation}
\mathcal{G}_{j}(t)=\left \langle \psi_{j,\uparrow}(0,t)\psi_{j,\uparrow}^{\dagger}(0,0)\right \rangle=\left \langle \psi_{j,\downarrow}(0,t)\psi_{j,\downarrow}^{\dagger}(0,0)\right \rangle\, ,
\end{equation}
the electron Green's functions on the edges ($j=A,B$) and in the antidot ($j=ad$) computed for the system in thermal equilibrium at temperature $T$. In performing the thermal averages, denoted by the brackets $\left\langle\ldots\right\rangle$, we assume that collective excitations are relaxed by extrinsic damping processes over a time-scale much shorter than the average electron dwell time through the system.~\cite{CavalierePRL04} As a consequence, the distribution $\left\{m_{\nu,i}\right\}$ is weighted with a Boltzmann thermal factor and a sum over all final collective plasmon states $\left\{m_{\nu,f}\right\}$ is then performed. A given process is completely specified by the initial quantum numbers of the antidot ($n,s$) and by their variation after the process ($\Delta_n, \Delta_s$) only, $\Gamma_{i\to f}^{(\lambda)}\to \Gamma^{(\lambda)}_{\Delta_n,\Delta_s}(n,s)$.\\

\noindent The tunneling rates can be recast in a more compact form
\begin{equation}\label{rateT}
\Gamma^{(\lambda)}_{\Delta_n,\Delta_s}(n,s)=\Gamma_0^{(\lambda)}\sum_{p=-\infty}^{\infty}a_p\gamma\left [-\Delta U^{(\lambda)}_{\Delta n, \Delta s}(n,s)-p\epsilon\right ],
\end{equation}
where, provided that $k_BT\ll\epsilon$, one has~\cite{Braggio00}
\begin{equation}\label{ap3}
a_p=\Theta(p+0^+)\frac{{\bf{\Gamma}} (\zeta+p)}{p!}, \ \ \ \ \ p\in\mathbb{Z}
\end{equation}
\begin{equation}\label{gamma}
\gamma(E)=\left (\frac{\epsilon}{\omega_c}\right )^{\zeta}\frac{e^{\frac{\beta E}{2}}}{{\bf{\Gamma}}(\zeta)}\left (\frac{2\pi}{\beta\omega_c}\right )^{\zeta-1} \mathcal{B}\left [\frac{\zeta}{2}-i\frac{\beta E}{2\pi},\frac{\zeta}{2}+i\frac{\beta E}{2\pi}\right ]
\end{equation}
with $\mathcal{B}$ the Euler Beta function, $\bf{\Gamma}$ the Euler Gamma function and $\Theta$ the Heaviside step function.
In Eq. ~\eqref{rateT} we introduced the energy cut-off $\omega_c= v_F/(Ka)$, the interaction dependent coefficients $\zeta=\left (K+1/K\right )/2$, the inverse temperature $\beta=1/k_BT$ and the characteristic rate
\begin{equation}
\Gamma_0^{(\lambda)}=\frac{|t_{\lambda}|^2}{(2\pi a)^2}\frac{1}{\omega_c}
,\end{equation}
on the barrier $\lambda$.
We point to Ref. \onlinecite{CavalierePRL04, Braggio00} for a detailed derivation of Eq. ~\eqref{rateT}.\\
\noindent In the sequential regime, the transport through the antidot can be approached by a master equation formalism.~\cite{Beenakker91, Blum} The occupation probabilities of the states $|n,s\rangle$ of the antidot in the stationary regime satisfy
\begin{eqnarray}\label{master}
&&\sum_{\lambda=A,B}\sum_{\Delta_n,\Delta_s=\pm 1}\left [
P_{n,s}\Gamma^{(\lambda)}_{\Delta_n,\Delta_s}(n,s)-\right .\nonumber \\
&&\left .P_{n+\Delta_n,s+\Delta_s}\Gamma^{(\lambda)}_{-\Delta_n,-\Delta_s}(n+\Delta_n,s+\Delta_s)\right ]=0
\end{eqnarray}
together with the normalization condition $\sum_{n,s}P_{n,s}=1$.
Charge and spin tunneling currents through the antidot between upper and lower edges are
\begin{equation}\label{Irho}
I^{(tun)}_{\rho}=-e\sum_{n,s,\Delta_n,\Delta s}\Delta_nP_{n,s}\left [\Gamma^{(A)}_{\Delta_n,\Delta s}(n,s)-\Gamma^{(B)}_{\Delta_n,\Delta s}(n,s)\right ]
\end{equation}
\begin{equation}\label{Isigma}
I^{(tun)}_{\sigma}=\frac{\hbar}{2}\sum_{n,s,\Delta n,\Delta_s}\Delta_sP_{n,s}\left [\Gamma^{(A)}_{\Delta n,\Delta_s}(n,s)-\Gamma^{(B)}_{\Delta n,\Delta_s}(n,s)\right ].
\end{equation}
\noindent The peculiar helical nature of the system reflects in a close connection between these tunneling currents and the {\emph{longitudinal}} currents flowing along the two edges of the bar.
First of all, tunneling processes are responsible for a decrease in the total charge current flowing from left to right contact with respect to the uncoupled edge configuration, i.e. $I=\frac{2e^2}{h}V-I_{\mathcal{BS}}$.
It has been shown~\cite{Strom09, Dolcetto12} that the backscattering current $I_{\mathcal{BS}}$ is directly related to the above spin tunneling current from edge to edge by
\begin{equation}\label{Ibs}
I_{\mathcal{BS}}=\frac{2e}{\hbar}I^{(tun)}_{\sigma}.
\end{equation}
The spin-polarized current flowing along the edges of the QSH bar~\cite{Citro11}
$I_{sp}=\frac{\hbar}{2}\sum_{\sigma=\uparrow, \downarrow}\left [\dot{n}_{A,\sigma}-\dot{n}_{B,\sigma}\right ]$ is zero in the uncoupled edge configuration.
However, when the antidot connects the two edges, a non-zero charge tunneling current is responsible for the generation of a longitudinal spin-polarized current
\begin{equation}\label{Isp}
I_{sp}=\frac{\hbar}{2e}I^{(tun)}_{\rho}
.\end{equation}
Ordinary two-terminal tunneling geometries, realized by means of QPCs or extended contacts, are not able to generate a spin-polarized current along the edges of the bar only by electrical means.~\cite{Strom09, Dolcetto12, Ferraro12}\\
\noindent However, what we are going to show is that, under certain conditions, the antidot acts as a {\em spin filter} on the longitudinal current, allowing to generate a spin polarization and to control it in a very simple and efficient way.

\section{Spin-polarized current}\label{spin-polarized current}
\begin{figure}[!ht]
\centering
\includegraphics[width=7cm,keepaspectratio]{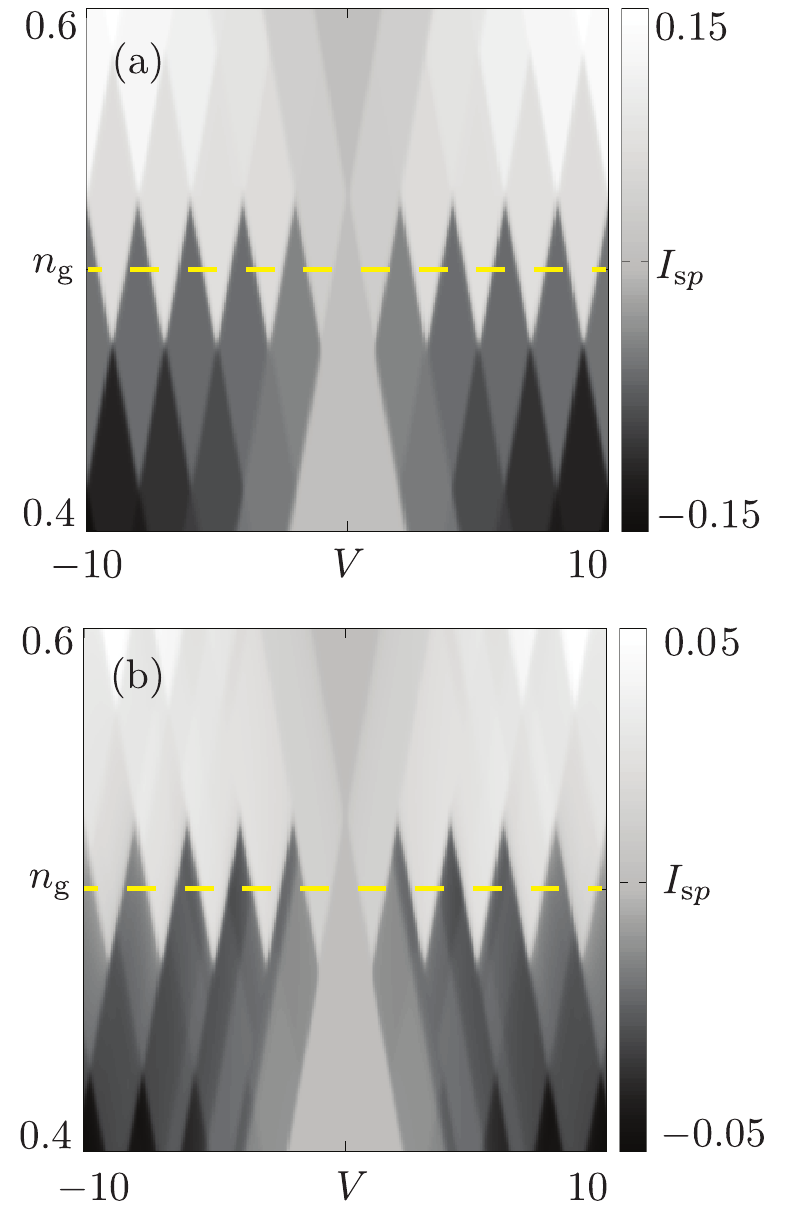}
\caption{(Color online) Spin-polarized current $I_{\mathrm{sp}}$ (units $\hbar\Gamma_{0}/2$) flowing along the edges of the bar as a function of $V$ (units $\epsilon_0/e$) and $n_{\rm{g}}$ for (a) $K=1$ and (b) $K=0.57$.
In both panels $\eta=10$, $k_{B}T/\epsilon_0=0.02 \epsilon_0$, $E_n/\epsilon_0=20$, $\omega_c/\epsilon_0=100$ and $n=0$.}
\label{density}
\end{figure}

\begin{figure}[!ht]
\centering
\includegraphics[width=7cm,keepaspectratio]{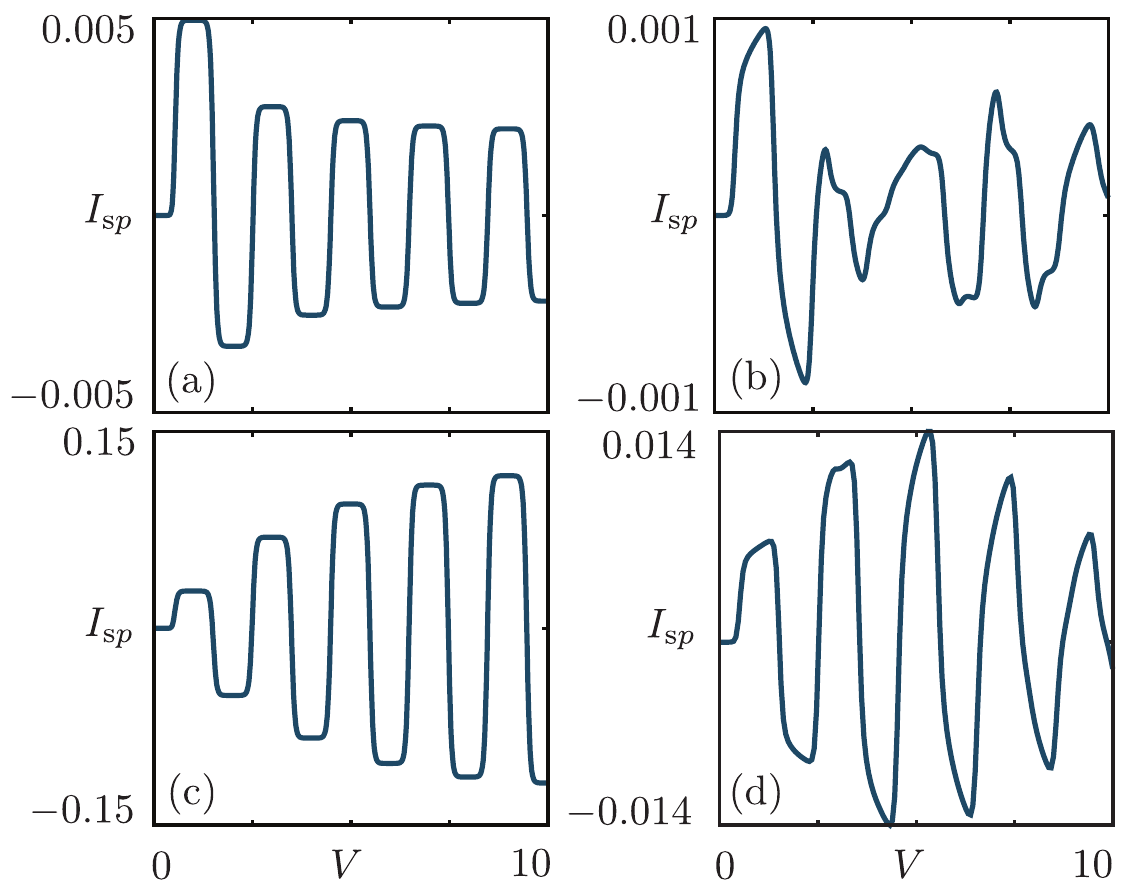}
\caption{(Color online) Plot of $I_{sp}$ (units of $\hbar\Gamma_0/2$) as a function of $V$ (units of $\epsilon_0/e$) for $n_{\rm{g}}=0.5$ with (a-c) $K=1$ and (b-d) $K=0.45$.
The asymmetry $\eta=1.5$ for the upper panels (a-b) and $\eta=20$ for the lower panels (c-d).
In all panels $k_{B}T/\epsilon_0=0.02$, $E_n/\epsilon_0=20$, $\omega_c/\epsilon_0=100$ and $n=0$.}
\label{scan_eta}
\end{figure}

At low temperature ($k_{B}T\ll\epsilon_{0}, eV$), the transport occurs predominantly via spin up (down) electrons jumping into the antidot through barrier $A$ ($B$), while spin down (up) electrons jump outside the antidot through barrier $A$ ($B$).
\begin{figure}[!ht]
\centering
\includegraphics[scale=0.6]{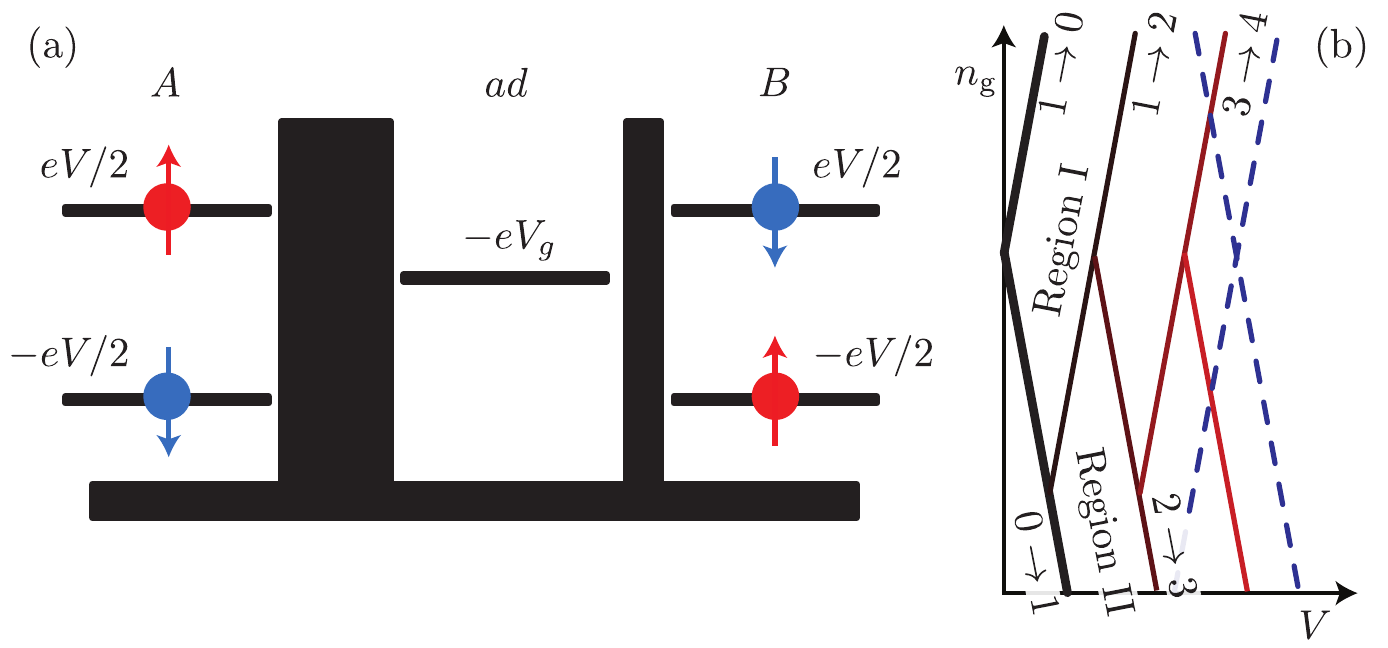}
\caption{(Color online) (a) Schematic representation of the antidot coupled to the edges of the bar through tunneling barriers, according to Fig.~\ref{antidot}. The bias voltage $V$
controls the chemical potential difference of the edges. The gate
voltage $V_g$ controls the chemical potential of the antidot.
(b) Scheme of the states involved in the transport. Thick black lines denote ground-to-ground state transitions, colored solid lines represent transitions involving zero-mode spin excited states, while dashed lines correspond to a transition involving a plasmon excitations.
The modulus of the initial ($I$) and final ($F$) spin states are denoted as  $|s_{I}|\to|s_{F}|$ (spin states $\pm s$ are degenerate in our model).}\label{schema_doppio}
\end{figure}
In Fig. \ref{schema_doppio}(a) the two tunneling barriers are represented with different thicknesses. We define here the unit of tunneling rate as $\Gamma_0\equiv \Gamma^{(A)}_0$ and introduce an asymmetry parameter $\eta$ so that $\Gamma^{(B)}_0=\eta\Gamma_0$.\\
\noindent Figure~\ref{schema_doppio}(b) depicts a scheme of different transport regions.
Coulomb blockade regions appear at low bias $V$, where sequential tunneling is blocked due to energy conservation constraints.
Tunneling becomes allowed at finite $V$ when $\Delta U^{(\lambda)}_{\Delta_n,\Delta_s}(n,s)<0$.
Moving out of the Coulomb blockade region, the lowest-lying transitions are those involving the ground states of $n$ and $n+1$ electrons (we assume here $n$ even for definiteness). The processes that load one extra electron into the antidot are $|n,0\rangle\to |n+1,+1\rangle$ through barrier $A$ and $|n,0\rangle\to |n+1,-1\rangle$ through barrier $B$, while the ones unloading it are $|n+1,1\rangle\to|n,0\rangle$ through barrier $B$ and $|n+1,-1\rangle\to|n,0\rangle$ through barrier $A$.
Those are represented as thick solid lines in Fig. \ref{schema_doppio}(b).
Increasing $V$, antidot excited states become also involved.\\
\noindent In the symmetric case ($\eta=1$), from Eqs. ~\eqref{DeltaU}, ~\eqref{mu} and ~\eqref{rateT} follows that $\Gamma^{(A)}_{\Delta_n,\Delta_s}(n,s)=\Gamma^{(B)}_{\Delta_n,-\Delta_s} (n,-s)$.
This property, together with Eqs. ~\eqref{Irho}, ~\eqref{Isp}, implies that $I^{(tun)}_{\rho}=0=I_{sp}$ at any $V$, so that longitudinal spin-polarized current along the edges {\em cannot} be generated.\\ \\
\noindent The situation with {\em asymmetric} tunnel barriers is however markedly different.
In the rest of the paper we consider the case $\eta>1$, the situation with $\eta<1$ being completely analogous swapping edge $A$ with $B$ and spin up with spin down.
Figure~\ref{density} shows the spin-polarized current $I_{sp}$ for (a) non-interacting ($K=1$) and (b) interacting ($K=0.57$) electrons.
Note that in the non-interacting case, collective excitations lines overlap with the ones involving zero-mode spin excited states.
In the interacting case, transitions are smoothed, because the edges of the bar are no longer Fermi liquid-like.
Furthermore, plasmon excitations are pushed to higher energies and, in general, do not overlap with the ones involving zero-mode spin excited states.
Note that two transport regions can be identified, supporting either positive or negative spin-polarized current.\\
\begin{figure}[!ht]
\centering
\includegraphics[width=7cm,keepaspectratio]{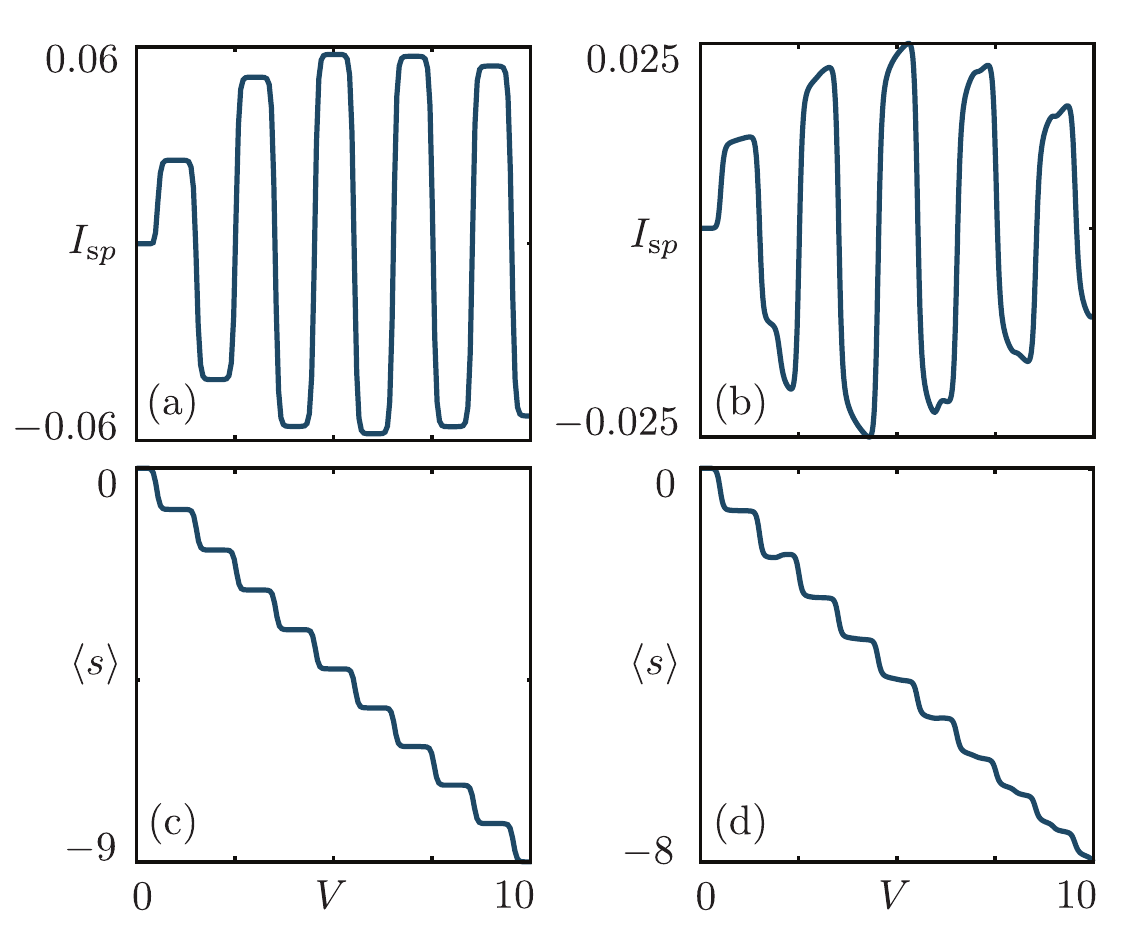}
\caption{(Color online) Plot of $I_{sp}$ (units of $\hbar\Gamma_0/2$, upper panels) and $\langle s\rangle$ (units of $\hbar/2$, lower panels) as a function of $V$ (units of $\epsilon_0/e$) for $n_{\rm{g}}=0.5$, corresponding to the dashed yellow line of Fig. \ref{density}, with (a-c) $K=1$ and (b-d) $K=0.57$.
As $V$ sweeps, $I_{sp}$ changes sign and $|\langle s\rangle|$ increases.
In all panels $\eta=10$, $k_{B}T/\epsilon_0=0.02$, $E_n/\epsilon_0=20$, $\omega_c/\epsilon_0=100$ and $n=0$.}
\label{taglio}
\end{figure}
\noindent Figure \ref{taglio} shows $I_{\mathrm sp}(V)$ for $n_{\rm{g}}=0.5$, for (a) non-interacting ($K=1$) and (b) interacting ($K=0.57$) electrons.
Differences in shape between the two panels are due to the effects of interactions discussed above.
However, by varying bias, in both cases the spin-polarized current alternates between positive and negative values.
The alternance in sign of $I_{\mathrm sp}$ is accompanied by a decrease in the average value of the spin of the antidot $\langle s\rangle=\hbar/2\sum_{n,s}sP_{n,s}$, as shown in Fig. \ref{taglio}(c) and (d). As we are going to show, these two effects are strictly connected.\\
\noindent A qualitative understanding of this mechanism can be captured by looking at the most relevant processes occurring in the system.
Let us focus the attention on regions $I$ and $II$, which are indicated in Fig. \ref{schema_doppio}(a): in region $I$ ($II$), the maximum spin state available is $|s|=1$ ($|s|=2$).
\begin{figure}[!ht]
\centering
\includegraphics[scale=0.34]{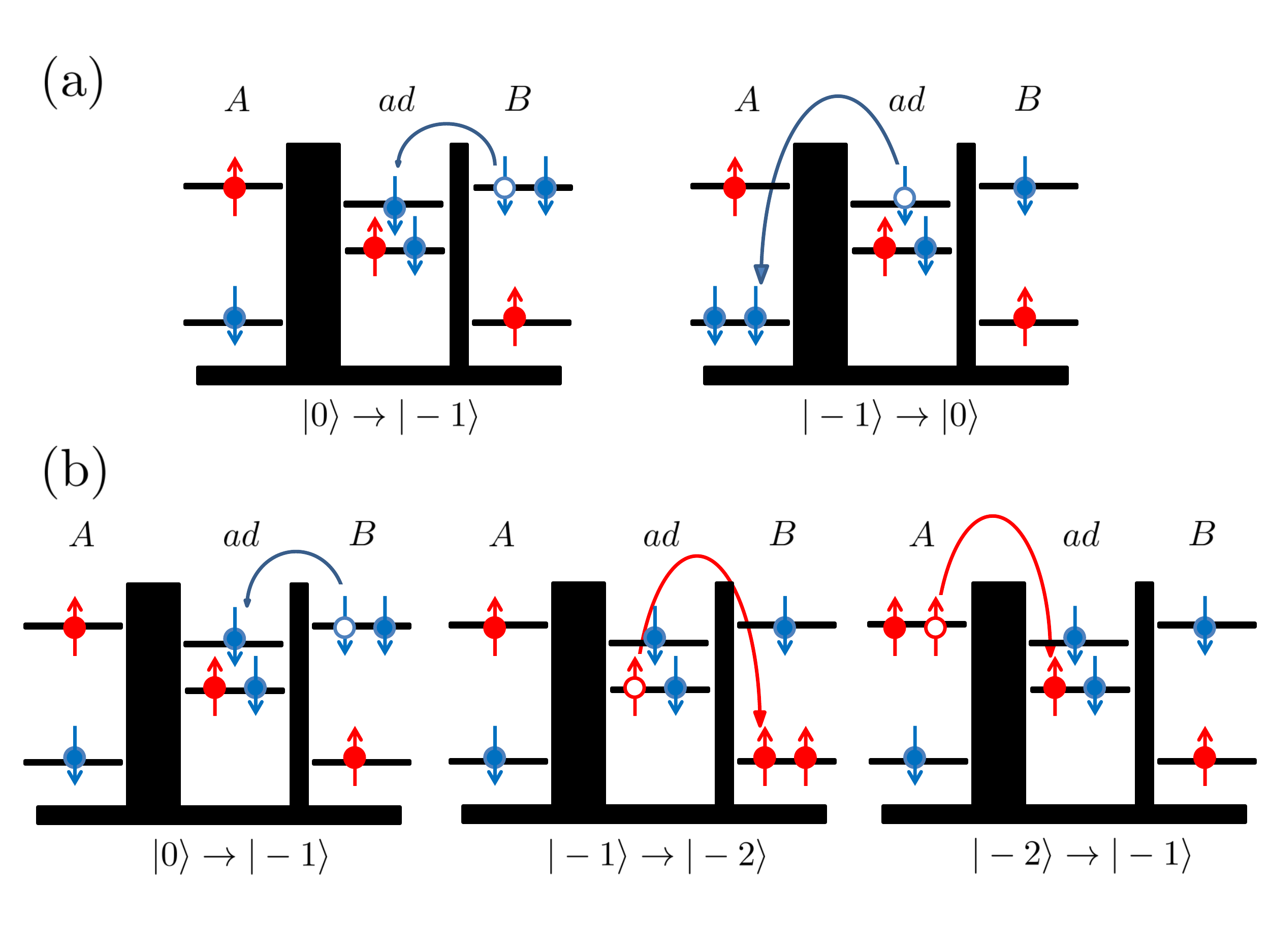}
\caption{(Color online) Scheme of the most relevant transport processes occurring in (a) region $I$ and (b) region $II$ (for $\eta>1$). See text for details.}\label{few_states}
\end{figure}
Consider region $I$ and the antidot in its ground state with an even number $n$ of electrons. The most likely scenario is a spin down electron leaking through barrier $B$ into the antidot, since spin up electrons from edge $A$ experience a thicker tunneling barrier. The spin down electron can now only escape through barrier $A$, and the sequence starts over.
The antidot most likely lies in the occupied state with $s=-1$.
These processes are depicted in Fig.~\ref{few_states}(a).
This mechanism implies a net flux of spin down electrons traversing the antidot from $B$ to $A$, leading to a net charge transfer and enabling the antidot to operate as a spin filter.
\noindent Tuning the antidot into region $II$, additional tunneling processes are allowed (see Fig. \ref{few_states}(b)). In particular, once a spin down electron tunnels into the antidot through $B$, a {\em spin up} electron can leave the antidot through the same barrier. This process, which leads the antidot into a state with $s=-2$, was {\em forbidden} in region $I$, due to energy constrains. The antidot is now "trapped" into the state with $s=-2$ until a spin up electron leaks through barrier $A$ into the antidot, leading back to the state with $s=-1$, and the sequence starts over. Thus, in this regime, a net flow of charge from edge $A$ to edge $B$ is created, which leads to a reverse of the current with respect to region $I$.
The antidot most likely lies in the unoccupied state with $s=-2$, decreasing the mean value of the spin of the antidot.
This mechanism is repeated when new zero-mode spin excited states become allowed.\\
\noindent In general, the value of $I_{sp}$ depends on asymmetry and interactions, as shown in Fig. \ref{scan_eta}, where the spin-polarized current is plotted for different values of $\eta$ and $K$.
In order to explain the detailed form of $I_{sp}$, plasmon excitations must be taken into account.
By increasing the bias, more and more collective excitations become allowed, eventually making it easier to tunnel across barrier $A$ than across $B$, even if the former is, in principle, thicker.
The competition between the asymmetry and the presence of collective excitations affects the value of the current.
From Fig. \ref{scan_eta}, we note that $|I_{sp}|$ develops a maximum for a certain value of bias.
If the asymmetry is little, the maximum is at low bias; by increasing the asymmetry, the maximum moves to higher bias.
\\
\\
\noindent In order to give quantitative analytical results on the behavior of $I_{\mathrm sp}$, we have developed a few-states model for regions $I$ and $II$, for the non-interacting case. In this regime, the allowed transitions are $|n,0\rangle\leftrightarrow|n+1,\pm 1\rangle$ (in both regions) and $|n+1,\pm 1\rangle\leftrightarrow|n,\pm 2\rangle$ (in region $II$ only).
The tunneling rates (units $\Gamma_{0}$) are given by
\begin{eqnarray}\label{rates}
\Gamma^{(A)}_{+1,+1}(0,0)&=&a_0 \ \ \ \ \ \ \ \ \ \ \ \ \ \ \Gamma^{(B)}_{+1,-1}(0,0)=\eta a_0 \nonumber \\
\Gamma^{(A)}_{+1,+1}(0,-2)&=&a_0+a_1 \ \ \ \ \ \Gamma^{(B)}_{-1,-1}(1,+1)=\eta(a_0+x a_1)\nonumber \\
\Gamma^{(A)}_{-1,+1}(1,-1)&=&a_0+x a_1  \ \ \ \Gamma^{(B)}_{+1,-1}(0,+2)=\eta (a_0+a_1)\nonumber \\
\Gamma^{(A)}_{-1,+1}(1,+1)&=&x a_0 \ \ \ \ \ \ \ \ \ \ \Gamma^{(B)}_{-1,-1}(1,-1)=\eta xa_0
.\end{eqnarray}
Here, $x=0$ if only states up to $|s|=1$ are involved in transport, while $x=1$ if states up to $|s|=2$ are.
From Eq. ~\eqref{ap3}, for $K=1$ one has $a_0=a_1=1$.
By solving the master equation one obtains the stationary spin-polarized current and the mean value of the spin of the antidot
\begin{equation}\label{I_sigma_5}
\frac{I_{sp}}{\hbar\Gamma_0}=\begin{cases} \frac{\eta(\eta-1)}{\eta^2+\eta+1} & \mbox{region }I \\
-\frac{2\eta(\eta-1)(\eta^2-\eta+1)}{(\eta^2+1)^2+2\eta(\eta^2+\eta+1)} & \mbox{region }II
.\end{cases}
\end{equation}
\begin{equation}\label{Smedio}
\frac{\langle s\rangle}{\hbar/2}=\begin{cases} -\frac{\eta^2-1}{\eta^2+\eta+1}\underset{\eta\gg 1}{\to} -1 & \mbox{region }I \\
-\frac{2(\eta^2-1)(\eta^2+\eta+1)}{(\eta^2+1)^2+2\eta(\eta^2+\eta+1)}\underset{\eta\gg 1}{\to} -2 & \mbox{region }II
.\end{cases}
\end{equation}
\noindent As expected, for $\eta=1$ no spin-polarized current is produced. At $\eta\neq 1$, spin-polarized current changes sign in passing from region $I$ to region $II$. Thus, the system behaves as an electrically controlled source of spin currents, whose sign can be reversed by tuning the bias.
Note that the value of $I_{sp}$ depends on $\eta$: for example, for $\eta<\eta^*\sim 2.9$, $|I_{sp}|$ in region $I$ is greater than $|I_{sp}|$ in region $II$, while the opposite happens if $\eta>\eta^*$.
This reproduces a maximum of $|I_{sp}|$ moving to higher bias by increasing the asymmetry, in agreement with Fig. \ref{scan_eta}.
As already observed, the switching of the spin-polarized current is related to a decrease of the mean value of the spin of the antidot.
\\
\\
\noindent These results lead to two important conclusions, namely that ($i$) an asymmetric antidot ($\eta\neq 1$) acts as a {\em spin filter} on the systems, leading to a {\em spin-polarized} longitudinal current. In addition, ($ii$) the sign of this spin-polarized current can be easily switched by tuning the bias voltage across the system.
Enhancement and/or stability in the presence of inter-edge interactions with self-consistent screening \cite{Sanchez13} deserve further investigations.

\section{Conclusions}\label{conclusions}
We proposed a device for generating and controlling spin-polarized currents in a two-terminal setup, without the need of magnetic fields or ferromagnets. The close connection between spin and chirality, hallmark of a quantum spin-Hall system, is the necessary condition to generate this process.\\
\noindent The spin-filter is realized in a two-terminal antidot configuration with asymmetrical tunnel barriers, in which helical edge states appear along the edges of the bar and around the antidot. By employing a master equation approach, appropriate for sequential tunneling regime, we have shown how the spin-filtering mechanism occurs, and we have related it to a population of the high excited antidot spin states. We have discussed the role of Coulomb interactions and asymmetry, showing how the spin-polarized current can be controlled by simple electrical means.\\
\section*{Acknowledgements}
We thank A. Braggio for useful discussions. The support of CNR STM 2010 program, EU-FP7 via Grant
No. ITN-2008-234970 NANOCTM, CNR-SPIN via Seed Project PGESE001 and MIUR-FIRB - Futuro in Ricerca 2012 - Project HybridNanoDev (Grant  No.RBFR1236VV) is acknowledged.

\end{document}